\documentclass[12pt]{article}
\usepackage[english,activeacute]{babel}
\usepackage{natbib}
\usepackage{comment}
\usepackage{float}
\usepackage[hidelinks]{hyperref}
\usepackage{mathrsfs}
\usepackage{enumitem}
\usepackage[font={small,it}]{caption}
\usepackage{amsmath,amsfonts,amsthm,amssymb}
\usepackage{bm,rotating,multirow,dsfont,graphicx}
\usepackage[usenames,dvipsnames]{color}
\usepackage{url}
\usepackage{multicol}
\usepackage{multirow}
\usepackage[T1]{fontenc}
\usepackage{flafter}
\usepackage{appendix}
\usepackage{subfigure}
\usepackage{xcolor}
\usepackage{soul}
\usepackage{threeparttable}
\makeatletter
\def\hlinewd#1{%
	\noalign{\ifnum0=`}\fi\hrule \@height #1 %
	\futurelet\reserved@a\@xhline}
\makeatother
\newcommand{\be}{\begin{eqnarray*}}
\newcommand{\ee}{\end{eqnarray*}}
\newcommand{\bet}{\begin{eqnarray}}
\newcommand{\eet}{\end{eqnarray}}
\def\spacingset#1{\renewcommand{\baselinestretch}{#1}\small\normalsize}\spacingset{1}
\def\@roman#1{\romannumeral #1}
\graphicspath{{./plots/}}

\begin{document}

\title{International Trade Network: \\Statistical Analysis and Modeling}

\author{
    Juan Sosa$^{1}$\footnote{Email: jcsosam@unal.edu.co} \\
    Andrés Felipe Arévalo-Arévalo$^{1}$ \\
    Juan Pablo Torres-Clavijo$^{1}$\\
}

\date{
    $^{1}$Universidad Nacional de Colombia, Bogotá, Colombia\\%
}

\maketitle

%\vspace{3cm}

\begin{abstract} 
\noindent
Globalization has rapidly advanced but exposed countries to supply chain disruptions, highlighted by the COVID-19 pandemic. This study exhaustively analyzes bilateral export data for 186 countries from 2018, 2020, and 2022, using Exponential Random Graph Models (ERGMs), to identify determinants of trade relationships, as well as Stochastic Block Models (SBMs), to characterize countries' roles in the trade network. Our findings show persistent, significant nodal characteristics driving bilateral trade and reveal no major structural changes in the trade network due to the pandemic.
\end{abstract}


\spacingset{1.1} % DON'T change the spacing!

\section{Introduction}\label{intro}

Over the past few decades, the world has witnessed a marked process of globalization, resulting in an unprecedented level of interconnection between countries \citep[e.g.,][]{Giddens2000,Stiglitz2002,Friedman2005}. 
Globalization has profoundly impacted the dynamics of the global economy, leading to a trend of liberalization, where technological and regulatory developments have promoted the free movement of capital and goods \citep[][]{Piketty2014,Rodrik2018,WorldBank2020}.

Similarly, globalization has transformed global production dynamics, as countries have chosen to specialize in the production of goods where they hold comparative advantages. 
These goods are then offered in the international market, while the production of items where they lack such advantages is outsourced to other countries. However, despite the benefits of trade dynamics fostered by globalization, this has introduced the risk of disruption in global supply chains. 
In other words, events that impede the normal flow of commercial activities can lead to shortages of goods and services, with no available alternatives or substitutes \citep{no_10_Covid}.

The risk of disruption to global supply chains materialized during the COVID-19 pandemic, a low-probability but high-impact event that disrupted international trade \citep{Petropoulos2021}. 
The World Health Organization (WHO) declared COVID-19 a pandemic on March 11, 2020, and it became one of the most profound crises of our time due to its vast socioeconomic impact. 
To curb the spread of the virus, many countries imposed various restrictions on individuals and businesses, including mandatory lockdowns and bans on foreign entry. 
According to \cite{no_11_Covid}, the strictest restrictions occurred during the first wave of the pandemic, which took place in the second quarter of 2020. However, during the second wave in the third quarter of 2020, countries began to gradually lift these restrictions, leading to a slow recovery in growth and a gradual reactivation of global economic activity.

As a result of the severe disruption to global supply chains and the product shortages caused by the COVID-19 pandemic, \cite{no_10_Covid} argue that several authors began to consider that the pandemic might signal the end of globalization or at least a permanent alteration of the process. 
To assess whether the COVID-19 pandemic had an impact on globalization processes, it is essential to analyze trade dynamics—from their determinants to their fundamental characteristics—before, during, and after the pandemic. Therefore, this article employs several network analysis methodologies to understand the short- and long-term implications of the pandemic. This approach aligns with \cite{estructura_red_no4}, who suggest that different economic crises have renewed interest in the study of trade networks, as economic interactions between countries are the primary channels through which economic shocks are propagated.

Several authors have analyzed a wide range of international trade network configurations to provide decision-making tools and inform policy development. Notable among them is \cite{Space_time_2012}, who studied trade networks in response to geographical and temporal changes, using diverse subnet configurations to demonstrate that the global trade network is disassortative at long distances but assortative at short distances. 
This contrasts with \cite{ml_red_no1}, who found that the network is generally disassortative. Another example is \cite{Trigo_2020}, who explored the global wheat trade network through an Exponential Random Graph Model, incorporating factors such as reciprocity, economic openness, country size, and land area. 
Their findings suggest that these factors are crucial in establishing trade links between countries. 
Additionally, \cite{Principal_2022} analyzed the global trade network for the years 2011, 2013, 2015, and 2017, using a broad range of ERGM configurations and characteristics. 
They concluded that positive effects of reciprocity and transitivity prevail, alongside the influence of non-structural factors in the formation of trade links.

It is also worth noting that the most popular methodologies used to study bilateral trade relations include: Gravity Equation Models for predicting links \citep[e.g.,][]{Eaton2002,Silva2006,Baier2009}, as employed by \cite{estructura_red_no4}, \cite{determinantes_no_7}, and \cite{no_11_Covid}; Fitness Models to replicate network topology \citep[e.g.,][]{Bianconi2001,Caldarelli2007,Miller2008}, as presented by \cite{estructura_red_no4}; and Exponential Random Graph Models to identify the determinants of bilateral trade relations \citep[e.g.,][]{Frank1986,Wasserman1996,Robins2007}, as carried out by \cite{Principal_2022} and \cite{Trigo_2020}.

This article characterizes the international trade network, from defining its topological properties to identifying the determinants of bilateral trade relations, as well as examining the role of different countries within the network. 
Additionally, it determines whether the COVID-19 pandemic led to short- or long-term changes in the network's structure. 
With this in mind, the study employs various methodologies for analyzing relational data, such as calculating and interpreting the structural characteristics of the network, using Exponential Random Graph Models (ERGMs; \citealt{Hunter2006,Snijders2010,Lospinoso2021}) as well as Stochastic Block Models (SBMs; \citealt{Holland1983,Nowicki2001,Karrer2011,Peixoto2014,Fortunato2016,Karrer2017}).

This article is structured as follows: The first section presents the data used for the analysis. 
The second section outlines the theoretical framework behind the methodologies and models employed in the study. 
The third section discusses the results of the analysis. 
Finally, the fourth section discusses the main findings and several lines for future research.

\section{International Trade Network: Construction and Processing}

We constructed the international trade network using two primary databases: One containing bilateral trade relations and the other detailing macroeconomic, geographic, and political characteristics of the countries under evaluation. 
In order to establish bilateral trade relations, we use data on the value of bilateral exports and imports reported annually in thousands of US dollars by the United Nations Statistics Division (UNSD). 
These data are stored in the World Integrated Trade Solution (WITS) repository. Data from the years 2018, 2020, and 2022 were extracted for 240 territories.

\cite{datos_faltantes_red_no2} argue that, in practice, measurements of exports and imports do not necessarily align due to methodological differences in data collection, which typically results in discrepancies. 
Therefore, export and import data should be analyzed separately. 
Consequently, in this article we exclusively focus on export data, primarily concentrating on exporting countries. 
This approach does not undermine the analysis, as export data should effectively represent the majority of global trade flows, since theoretically, exports from country A to country B should be equal to the imports of country B from country A \citep{BussierChinn2018}.

\begin{table}[!b]
\centering
\begin{threeparttable}
\caption{Descriptive statistics of the bilateral trade relations database.}
\label{Tabla_1}
\begin{tabular}{l|c|c|c}
    \hline
    Metric & 2018 & 2020 & 2022 \\
    \hline
    Number of countries & 186 & 186 & 186 \\
    Number of observations & 16,357 & 18,017 & 17,276 \\
    Average trade flow & 905,674 & 764,221 & 1,114,961 \\
    Median  trade flow & 7,054 & 5,482 & 8,052 \\
    Standard deviation & 7,415,086 & 6,532,794 & 9,699,058 \\
    Minimum & 0 & 0 & 0 \\
    Maximum & 479,278,747 & 452,492,876 & 582,756,110 \\
    \hline
\end{tabular}
\begin{tablenotes}
    \item Source: Prepared by the authors based on WITS data.
\end{tablenotes}
\end{threeparttable}
\end{table}

The bilateral trade relations database consists of 240 territories, including countries, special administrative regions, economic and monetary unions, among others. 
For this analysis, we chose countries as the unit of analysis.
Therefore, the list of ISO country codes recognized by the United Nations (UN) can be used to filter the original database accordingly. 
It is important to note that the WITS database does not provide reports for all countries, resulting in a final dataset comprising 186 countries. 
Table \ref{Tabla_1} presents the main descriptive statistics of the bilateral trade relations database. 
The data exhibit considerable dispersion, suggesting significant asymmetries between countries, as evidenced by the difference in magnitude between the median and mean of export data.

\begin{table}[!b]
\centering
\begin{threeparttable}
\caption{Descriptive Statistics of the node attribute database for 2018.  Agriculture, Industry, and Services are computed using added values. GDP and GCF stand for Gross Domestic Product and Gross Capital Formation, respectively. Total Population is given in millions. Landlocked is a dummy variable assuming the value 1 if the country has access to the sea, and 0 otherwise.}
\label{Tabla_2}
\small
\begin{tabular}{l @{\hspace{0.04cm}}|c|c|c|c|c @{\hspace{0.04cm}}}
    \hline
    Attribute & Mean & SD & Min. & Median & Max. \\
    \hline                      
    Agriculture (GDP \%) & 10.33 & 10.34 & 0.02 &  6.73 & 58.93 \\
    Industry (GDP \%)  &  26.12& 12.01 & 4.87 & 24.65 & 65.88\\
    Services (GDP \%) & 55.87 & 11.99 & 32.35 &  55.54 & 86.85 \\  
    GDP per capita (USD) & 16,236.00 & 27,658.80 & 232.10 & 6,316.5 & 193,968.10 \\ 
    GCF (GDP \%) & 24.61 & 7.94 &2.78 & 23.35 & 54.02\\
    Foreign investment (GDP \%) & -5.27  & 61.66 & -13.03 & 2.73 &29.21 \\
    Inflation (annual \%) & 4.96 & 14.21 & -2.89 &  2.79 & 55.97 \\
    Political stability & -0.06 & 0.97 & -2.75 & -0.01 & 1.58 \\    
    Education expenditure (GDP \%) &  4.37 & 1.84 &  0.23 & 4.37 &  15.38\\ 
    Total population & 38.30 &  150.49 & 0.01 & 7.86 &  1.40 \\
    Unemployment (workforce \%) & 7.39 & 5.87 & 0.11 &  5.20 &  26.98\\
    Landlocked &0.19 & 0.39& n/a & n/a & n/a \\
    \hline
\end{tabular}
    \begin{tablenotes}
        \centering
            \item Source: Prepared by the author based on World Bank data.
    \end{tablenotes}
\end{threeparttable}
\end{table}

In order to construct the node attribute database, we use macroeconomic, geographic, and political data for the countries under analysis. 
This information was sourced from the World Bank Development Indicators database (WB) and the Centre d’Études Prospectives et d’Informations Internationales (CEPII). 
The selection of variables was guided by economic theories concerning potential determinants of international trade.

Table \ref{Tabla_2} presents the main descriptive statistics of selected variables from the nodal attributes database. 
Each of the chosen nodal attributes represents a descriptive component of the characteristics of each country, ranging from the definition of comparative advantages to the description of political, economic, and social conditions. 
The variables 'Agriculture', 'Industry', and 'Services' are selected as proxies for the level of productive specialization in an economy. 
The variable 'Gross Domestic Product per capita' is chosen as a proxy for the level of wealth, and consequently, the potential consumption capacity of an economy. 
The variable 'Education Expenditure' is used as a proxy for human capital. The variables 'Gross Capital Formation' and 'Foreign Investment' are selected as proxies for infrastructure, technological advancement, and foreign investment. Lastly, the variables 'Unemployment', 'Population', 'Inflation', and 'Political Stability' are selected as proxies for the political and social situation of the economies.

\subsection{Handling Missing Data}

As in many studies involving macroeconomic data, the collected dataset contains gaps or missing values. 
In this case, we only consider variables with less than 30\% of missing data. 
To address this issue, the missing data values are imputed using the $k$-nearest neighbors method (KNN Imputer; \citealt{buuren2010mice,pedregosa2011scikit}). 
For each missing value in the dataset, the KNN algorithm locates the $k$ nearest neighbors based on Euclidean distance and imputes the missing values by leveraging the observed values of the most similar data points within the feature space \citep{VIM}.
We use a cross-validation approach to determine the value of $k$ that minimizes the mean squared error, which give us an optimal number of $k = 1$ neighbors.

\subsection{Network Consolidation}

We define the international trade network a graph generated from bilateral export data between countries obtained from the World Integrated Trade Solution (WITS) for a given time period \( t \). 
Specifically, the graph \( G_t \) is defined as \( G_t = (V_t, L_t) \), where \( V_t \) represents the set of countries (nodes or vertices) and \(L_t \) denotes the number of directed trade links (edges or links), forming the network \( N_t = (V_t, L_t, P_t, W_t) \). Here, \( P_t \) is the node attribute matrix and \( W_t \) is the matrix of link weights. 
This implies that the links have varying intensities based on the value of trade flows \citep{Space_time_2012}.

\begin{figure}[!htb]
    \centering
    \includegraphics[scale = 0.52]{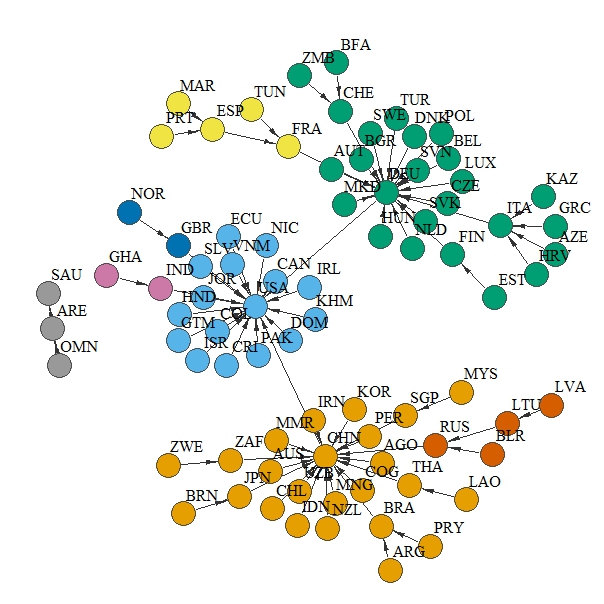}
    \caption{Graph for the 2018 international trade network filtered by the most notable countries of each class. Clusters are defined using the Infomap algorithm \citep{rosvall2008maps}.}
    \label{fig:grafo_top_80_2018}
\end{figure}

In general, each pair of nodes \( (i, j) \) is referred to as a dyad. For each dyad, the presence or absence of a link is indicated using the adjacency matrix, denoted as \( Y \). 
As mentioned above, the network is directed; thus, for any two nodes \( i \) and \( j \), \( Y_{i,j} = 1 \) (\( Y_{i,j} = 0 \)) represents the presence (absence) of a directed edge from node \( i \) to node \( j \). 
Specifically, \( Y_{ij} = 1 \) indicates that there is a trade relationship where country \( i \) exports to country \( j \). Additionally, the weight represents the magnitude of the exports, measured in thousands of US dollars.
Since the network is directed, \( Y \) is not necessarily symmetric \citep{bloques_1}.
Figure \ref{fig:grafo_top_80_2018} shows the graph for the 2018 international trade network filtered by the most notable countries of each class.

\section{Methodology}

In this section, we present the key methodological aspects for studying international trade networks. 
We primarily focus on a detailed characterization of Exponential Random Graph Models (ERGMs) and Stochastic Block Models (SBMs). For a gentle introduction to these topics, we recommend consulting the book by \citep{R_analisis_sosa}.

\subsection{ERGM}

Exponential Random Graph Models (ERGMs) serve as a direct extension of classical generalized linear models (GLMs; \citealt{McCullagh1989,Faraway2016,Hastie2017}) to the context of network data. 
The formulation of ERGMs is intended to adapt and extend the principles and methods of GLMs to the unique characteristics of network data.
While GLMs are used to model relationships between variables in traditional statistical settings, ERGMs are specifically designed to analyze the structure of networks by modeling the probability of observing a particular network configuration.

ERGMs, also known as $p$-star models, allow us to analyze link formation within networks. 
A key distinction from regression models is their ability to account for the interdependence of network links, relaxing one of the most restrictive assumptions inherent in regression models. 
This capability allows ERGMs to describe the fundamental forces shaping the global network structure. 
In this way, ERGMs enable the inclusion and consideration of node characteristics and structural configurations \citep{Principal_2022}. 
This makes them particularly suited for capturing the dependency relationships that define the global structure of international trade networks. 
That is why we use ERGMs to identify the principal factors influencing the formation of trade relationships between countries.

The general form of ERGMs specifies the probability of the entire network as a function of a set of relevant features.
Consider the random graph \( G = (V, E) \), where \( Y_{i,j} \) is a binary random variable indicating the presence or absence of an edge \( e \in E \) between two vertices \( i \) and \( j \) in \( V \). The basic specification for an ERGM is as follows,
\begin{equation*}
    \Pr(Y=y) = \frac{1}{\kappa} \exp\left(\theta_{1}g_{1}(y) + \theta_{2}g_{2}(y) + \ldots + \theta_{p}g_{p}(y)\right)\,,
\end{equation*}
where:
\begin{itemize}
    \item \( Y \) is the adjacency matrix.
    \item \( \kappa \) is the normalization constant.
    \item \( g_{h}(y) \) takes the value 1 if the edge configuration \( h \) is present in \( y \), and 0 otherwise.
    \item \( \theta_{h} \) is an unknown parameter indicating that dyads are dependent in the edge configuration \( h \).
\end{itemize}

It is important to note that, given a choice of functions \( g_{H} \) and their coefficients \( \theta_{H} \), there is a corresponding structure of dependence among the elements of \( Y \). 
In general terms, this structure can be described as specifying that the random variables \( \{Y_{ij}\}_{(i,j) \in A} \) are independent of \( \{Y_{i'j'}\}_{(i',j') \in B} \), conditional on the values of \( \{Y_{i'',j''}\}_{(i'',j'') \in C} \), for some given index sets \( A \), \( B \), and \( C \) \citep{R_analisis_sosa}.

The ERGMs framework is highly adaptable, offering various modifications and extensions suitable for different types of network data. 
It can be applied to both directed networks (such as trade networks) and undirected networks. 
Additionally, ERGMs enable the incorporation of vertex-specific information (such as economic indicators in trade networks), which enhances the model’s ability to capture node-level heterogeneity. 
When applied to the global trade network, the ERGM framework aims to identify significant parameters that accurately represent the network’s structure and relationships. By fitting the model to the trade data and analyzing these parameters, we are able to uncover underlying patterns and factors that influence trade interactions between countries, while providing valuable insights into the complexities of the global trade system.

In this study, we use the \texttt{statnet} package \citep{statnet} in \texttt{R} \citep{Rcore} to fit the Exponential Random Graph Model (ERGM). \texttt{statnet} offers comprehensive tools for network analysis, allowing us to specify and estimate ERGMs effectively.

\subsection{SBM} \label{section: Bloques_estocasticos}

Stochastic Block Models (SBMs)  are increasingly popular in statistical network analysis because they allow for the identification of latent group structures within a network \citep{bloques_1}.
By grouping nodes into blocks or clusters based on shared connection patterns, SBMs offer a powerful framework for detecting underlying communities or clusters. This approach is especially useful for simplifying complex networks and improving the efficiency of clustering tasks, as it reduces noise by focusing on the most meaningful interactions.

Following very closely \cite{bloques_1} and \cite{R_analisis_sosa}, in Stochastic Block Models (SBMs), each node is assigned to one of \( K < n\) unknown groups or clusters, where $n$ is the number of nodes. 
For each node \( i \), we define \( Z_i \) as a \( K \)-dimensional vector such that
$$
z_{i,p} = 
\begin{cases} 
1, & \text{if node $i$ belongs to community $p$;} \\
0, & \text{otherwise.}
\end{cases}
$$
By employing these \( K \)-dimensional vectors, we can construct a \( K \times K \) matrix \( C \) is such a way that each entry \( C_{p,q} \) represents the probability of a directed edge originating from a node in group \( p \) and terminating at a node in group \( q \). 
This matrix \( C \) is crucial for understanding and modeling how edges are formed within the network based on the group affiliations of the nodes. Specifically, \( C \) captures the likelihood of interactions between different groups, thereby providing insight into the underlying mechanisms of edge formation within the network. By analyzing this matrix, we can infer how the network’s structure is influenced by the block assignments and better understand the dynamics of connections across different groups.

Specifically, given \( Z \) and \( C \), the model can be expressed under the assumption that edges are distributed according to a Bernoulli distribution conditional on group memberships. For directed networks, the likelihood is given by
\begin{equation*}
p(Y \mid Z, C) = \prod_{i,j:i \neq j}(Z_i^{\textsf{T}} C Z_j)^{Y_{i,j}} (1 - Z_i^{\textsf{T}} C Z_j)^{1 - Y_{i,j}}\,.
\end{equation*}
However, in practical applications with real data, \( Z \) and \( C \) are not known (as well as the number of groups), and thus, must be inferred. 
To address this issue, we assume that \( \Pr(Z_{i,p} = 1\mid\theta_p ) = \theta_p \), where \( \sum_{p=1}^{K} \theta_p = 1 \). 
Therefore, given the values of $Z_1,\ldots,Z_n$, dyads can be modeled as conditionally independent according to the Bernoulli distribution:
$$
y_{i,j}\mid\mathbf{C},Z_i,Z_j \sim \textsf{Bernoulli}\left( C_{\xi_i,\xi_j} \right)\,,
$$
where $\xi_i = \xi(Z_i)$ denotes the position in $Z_i$ with a non-zero entry (i.e., $\xi_i = p$ indicates that node $i$ belongs to community $p$).

\cite{R_analisis_sosa} emphasize that estimating the SBM using maximum likelihood methods involves substantial computational costs, particularly due to the complexity of calculating the exact likelihood function for large networks. This computational burden makes it challenging to apply traditional methods, especially in cases with large-scale networks. To address these challenges, the authors suggest using the variational Expectation-Maximization (EM; \citealt{Beal2003,Wainwright2008,Blei2017}) algorithm as a more efficient alternative for model fitting. This algorithm circumvents the need to compute the full likelihood by optimizing a lower bound on the likelihood of the observed data, commonly referred to as the Evidence Lower Bound (ELBO). By doing so, the variational EM approach reduces the computational load and enables the estimation of the SBM in a more scalable and practical manner, while still maintaining reasonable accuracy in capturing the underlying community structure of the network.

\section{Results}

In this section, we present our key findings, including a comprehensive characterization of international trade networks for 2018, 2020, and 2022 using structural descriptive analysis, as well as full modeling analysis using ERGMs and SBMs. All results are fully reproducible. The corresponding code freely available at \url{https://github.com/aarevaloa/Red-de-comercio-mundial}.

\subsection{Network Structure}

Table \ref{Tabla_3} displays the main structural characteristics of the international trade networks for 2018, 2020, and 2022. 
From 2018 to 2020, the number of edges increased by 1,660 connections, while from 2020 to 2022, the number of edges increased decreased by 741. This suggests a shift in long-term economic dynamics in 2020, followed by a return to previous trends in 2022. This pattern is consistent with observations during the COVID-19 pandemic, where a significant acceleration in e-commerce growth was driven by strong export performance in East Asian economies \citep{UNCTAD}.

\begin{table}[!htb]
\centering
\begin{threeparttable}
\caption{Structural statistics of the international trade network. CV: Coefficient of variation (\%). Weight in millions of USD.}
\label{Tabla_3}
\small
\begin{tabular}{l|c|c|c}
    \hline
    Statistic & 2018 & 2020 & 2022 \\
    \hline                      
    Nodes (n°)       & 186 & 186 & 186 \\            
    Edges (n°)       & 16.357 & 18.017 & 17.276 \\
    Components (n°)  & 1.00 & 1.00 & 1.00 \\
    Mean out-degree  & 87.94 & 96.86 & 92.88 \\
    CV out-degree    & 85.01 & 73.35 & 78.97\\
    Mean in-degree   & 87.94 & 96.86 & 92.88 \\
    CV in-degree     & 0.24 & 0.25 & 0.25 \\
    Mean weight      & 905.67 &  764.22 & 1,114.96 \\ 
    CV weight        & 8.19 &  8.55 & 8.70 \\
    Degree correlation & 0.66 & 0.72 & 0.74 \\
    Density          & 0.47 & 0.52 & 0.50 \\
    Transitivity     & 0.76 & 0.79 & 0.78 \\
    Edge reciprocity & 0.62 & 0.69 & 0.66 \\
    Dyad reciprocity & 0.45 & 0.52 & 0.50 \\
    Associativity    & -0.23 & -0.24 & -0.24 \\
    \hline
\end{tabular}
\begin{tablenotes}
    \item Source: Prepared by the authors based on WITS data.
\end{tablenotes}
\end{threeparttable}
\end{table}

Although the pandemic led to a decrease in export values, this did not correspond to a reduction in the number edges. 
The economic, social, and health crises caused by COVID-19 initially did not result in the destruction of trade relationships but rather in a weakening of trade values, as reflected by the decline in the average edge weights in 2020. This weakening of trade ties prompted the formation of new trade relationships among agents who had not previously traded with each other, as part of efforts to secure essential supplies and food to address the health crisis and the needs of the population.
By 2022, we observe a recovery in export values, as indicated by the increase in the average edge weights, with levels surpassing those seen in the pre-pandemic period. Simultaneously, there is a reduction in the number of trade relationships, which can be attributed to the stabilization of production structures across economies.

There is a clear presence of a disassortative relations. 
Specifically, the assortativity coefficients for 2018, 2020, and 2022 are -0.23, -0.24, and -0.24, respectively. 
These negative values indicate a persistent trend where highly connected nodes tend to form associations with nodes that have substantialy fewer connections. Rather than forming clusters with similarly connected nodes, central or highly networked nodes preferentially interact with less connected, peripheral nodes. This finding aligns with prior research \citep[e.g.,][]{Space_time_2012,ml_red_no1}, which also identified a marked disassortative pattern in commercial relationships within the international trade network. 
Such a pattern suggests a distinctive structural dynamic in global trade, where key players engage with smaller or less prominent countries, reinforcing the hierarchical nature of these interactions. Such disassortativity has implications for understanding the flow of goods, influence, and resilience within the network.

The coefficients of variation exhibit consistently high values, which corresponds to a pronounced heterogeneity in how countries are integrated into the international trade network. 
This substantial variability suggests that countries do not participate in global trade on equal terms; instead, nations assume predefined roles within the network of trade relations. 
These roles may be influenced by various factors such as economic size, geopolitical position, trade agreements, or the nature of their export and import markets. 
The elevated coefficients underscore the unequal distribution of trade connections, where some countries act as central hubs with extensive trade links, while others maintain more peripheral roles with fewer connections. 
This structural disparity highlights the complexity exhibited by global trade, where the dynamics of influence, access to markets, and economic interdependence vary greatly from one country to another.

The international trade network is characterized by a high level of connectivity. 
This is evident from the network's density, which hovers around 0.5 for all three years.
Such high connectivity is a key indicator of the network’s global cohesion and it constitutes a likely explanation for the network's transitivity. 
The underlying reason for this pattern can often be found in the dynamics of trade treaties and the creation of regional trade blocs. Once two countries establish a commercial relationship, they are more likely to enter into multilateral agreements or participate in regional trade organizations that encourage broader interaction. As such, the high density of the network reflects not only the individual trade agreements between pairs of countries but also the broader institutional structures that facilitate global trade, such as free trade agreements, customs unions, and economic partnerships.

Finally, the analysis shows that the network remains a single connected component across all three years. This indicates that, despite geopolitical tensions and geographic barriers, global trade is effectively maintained, allowing for the exchange of goods and services between any pair of countries.
The presence of a single component highlights the global nature of trade, where no country is fully isolated from the system. Despite regional conflicts, trade disputes, or other obstacles that may create friction in international relations, these findings suggest that such challenges do not fragment the overall network. Instead, countries remain interconnected through various trade routes, often facilitated by intermediaries or regional trade agreements, ensuring that global commerce continues to operate as a cohesive system.

\subsection{Network Connectivity}\label{section: Conectividad}

To explore the connectivity of the global trade network, we follow very closely the reasoning provided in \cite{Space_time_2012}. 
Specifically, we partition the network based on net export values to create subgraphs with thresholds of export values. 
This approach allows us to analyze how the network's connectivity evolves as we introduce higher-weight edges. 
In this way, We examine changes in connectivity through metrics such as the number of connected components and the size of the largest connected component.

\begin{figure}[H]
\centering
\caption{Connectivity of the 2018 international trade network conditioned by export values. The blue dashed line marks the threshold beyond which the size of the giant component begins to decrease.} \label{fig:grafico_1}
\begin{threeparttable}
\includegraphics{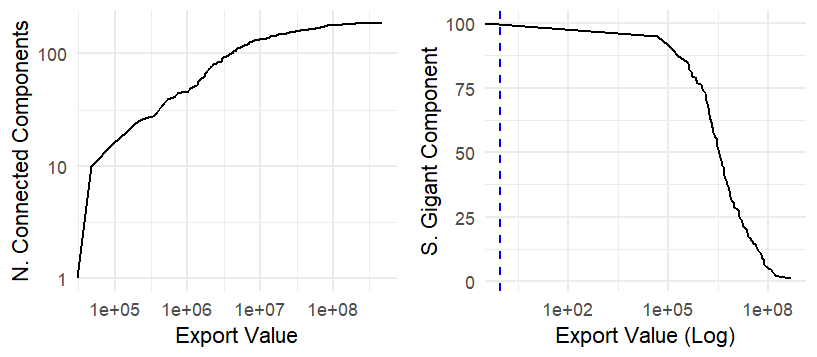}
\begin{tablenotes}
\centering
\item Source: Created by the authors based on data from WITS.
\end{tablenotes}
\end{threeparttable}
\end{figure}

Figure \ref{fig:grafico_1} illustrates how connectivity statistics change with accumulated export values for 2018 (similar connectivity graphs for 2020 and 2022 are not shown here since they are quite similar). Generally, as higher-value trade relationships are added to the network, the number of connected components increases. 
Initially, when only trade relationships with the lowest export values are included, the network shows a rapid rise in the number of connected components.
As trade relationships with higher export values are progressively added, the network's connectivity improves, leading to an increased number of connected components, enhancing the integration and cohesion of the network.

The network remains cohesive primarily due to numerous lower-value connections, which are crucial for maintaining overall connectivity. When these weaker connections are removed, the network fragments rapidly, revealing that only a few countries with significant high-value exports form its core. These core hubs anchor the network, while its stability depends on the extensive web of lower-value connections. The stabilization at higher values after the removal of these weaker links indicates that the remaining network comprises several smaller, densely interconnected groups of countries. This fragmentation also highlights that many nations outside the core rely heavily on these lower-value connections for their integration, becoming isolated when these links are removed.

The size of the giant component as a function of the logarithm of export values illustrate how the giant component’s size changes when only higher-value edges are considered. Initially, the giant component encompasses 100\% of the nodes, as all nodes are connected within a single component. However, as lower-value connections are removed, the size of the giant component begins to decrease but remains relatively stable until reaching a certain threshold (the inflection point of the graph). This indicates that the size of the giant component is primarily dependent on higher-value edges.

Our findings reflect the characteristics of the international trade network, where high-value connections are linked to large economies that export substantial quantities of goods and services, which are relatively rare due to the limited number of such economies. These large economies form central hubs in the network due to their extensive trade relationships. In contrast, smaller economies, with limited export capacity, rely on a few low-value connections and trade with fewer countries. The removal of these lower-value connections causes these smaller economies to gradually detach from the giant component. Conversely, when higher-value edges are removed, the central hubs (large economies) begin to disconnect, leading to the rapid disintegration of the giant component, highlighting the crucial role of these large economies in maintaining network cohesion.

\subsection{Exponential Random Graph Model}

This section presents the results regarding the ERGM, considering key non-structural characteristics influencing the international trade network. Notably, we identify several statistically significant effects on the formation of network edges. Table \ref{Tabla_4} shows the results for the 2018 network (similar tables for 2020 and 2022 are not shown here since they are quite similar).

Note that the covariates `nodecov agriculture', `nodecov foreign investment', `nodecov GDP per capita', and `nodecov education expenditure' are statistically significant with negative estimates. This result indicates that the probability of two countries establishing trade relations decreases as the presence of one or more of these characteristics weakens within their economies. Additionally, these four characteristics are established as fundamental descriptors of a country’s economic structure. Consequently, as two countries display divergent economic conditions, the likelihood of trade between them is expected to decline. This finding aligns with \cite{Comercio_caracteristicas}, which reported that dissimilar economic characteristics between countries adversely affect the formation of trade relationships.
Additionally, we observe the previous behavior persists over time. For both 2020 and 2022, the same covariates consistently show negative and statistically significant estimates.

\begin{table}[!htb]
\centering
\begin{threeparttable}
\caption{ERGM results for the 2018 network. Edges: Total number of edges in the graph. Nodecov: Sum of the attributes. Absdiff: Absolute difference of the attributes. Nodematch: Number of attribute matches. Significance: p value < 0.1 (.), p value < 0.05 (*), p value < 0.01 (**),  p value < 0.001 (***).}
\label{Tabla_4}
\small
\begin{tabular}{l|c|c|c|c}
    \hline
    Variable & Estimation & SD & z value & Significance \\
    \hline                      
    edges & -3.32 & 0.35 & -9.493 & *** \\            
    nodecov agriculture & -0.040 & 0.002 & -20.104 & ***\\
    nodecov industry & 0.010 & 0.002 & 5.127 & *** \\
    nodecov foreign investment & -0.001 & 0.000 & -4.815 & *** \\
    nodecov PIB per capita & 0.000 & 0.000 & -26.433 & *** \\
    nodecov education expenditure & -0.081 & 0.006 & 14.177 & *** \\
    nodecov GCF & 0.021 & 0.001 & 16.781 & *** \\
    nodecov inflation & 0.010 & 0.001 & 8.043 & *** \\
    nodecov political stability & 0.070 & 0.010 & 5.522 & *** \\
    nodecov total population & 0.000 & 0.000 & 7.733 & *** \\
    nodecov services & 0.030 & 0.002 & 13.270 & *** \\
    nodecov unemployment & 0.001 & 0.002 & 0.783 & \\
    nodematch agriculture & 0.080 & 0.500 & 0.153 & \\
    nodematch industry & -1.030 & 1.050 & -0.977 & \\
    nodematch foreign investment & -0.230 & 0.550 & -0.425 & \\
    nodematch PIB per capita & -10.010 & 70.020 & -0.143 & \\
    nodematch education expenditure & 0.510 & 0.240 & 2.080 & * \\ 
    nodematch GCF & -0.370 & 0.300 & -1.221 & \\
    nodematch inflation & -2.320 & 1.090 & -2.140 & *\\
    nodematch political stability & 1.340 & 1.080 & 1.241 \\
    nodematch total population & 9.380 & 70.020 & 0.134 & \\
    nodematch services & 1.220 & 1.230 & 0.991 & \\
    nodematch unemployment & -0.450 & 0.370 & -1.219 &\\
    nodefactor landlocked & -0.300 & 0.020 & -13.736 & *** \\
    \hline
\end{tabular}
\begin{tablenotes}
\centering
    \item Source: Prepared by the authors based on data from WITS and the World Bank.
\end{tablenotes}
\end{threeparttable}
\end{table}

The covariates `nodecov industry', `nodecov GCF', `nodecov inflation', `nodecov political stability', `nodecov total population', and `nodecov services' exhibit positive and statistically significant estimates, suggesting that these factors are associated with an increased likelihood of establishing trade links.
Independent evaluation of these covariates is consistent with economic literature. For example, `nodecov inflation' indicates that rising inflation often results in currency depreciation as agents move their assets to safer investments. This depreciation makes local goods cheaper internationally, increasing export demand. However, the effectiveness of raising interest rates to combat inflation depends on perfect arbitrage between countries; otherwise, its impact will vary based on factors such as expectations, institutional confidence, and foreign exchange market elasticity.

The covariate `nodecov inflation' is statistically significant in 2018 but it is not in 2020 (see Table \ref{Significancias_18_20_22}). This shift can be attributed to the pandemic-induced sharp decline in household and business consumption, which suppressed inflationary pressures and led to a global period of low inflation. As a result, the variable loses its significance as a determinant of trade between countries during this period. However, by 2022, inflation levels had reverted to pre-pandemic values due to economic recovery and increased consumption by households and businesses, which corresponds with the variable’s return to statistical significance.

When evaluating nodal characteristics by differences, only `nodematch education' expenditure' and `nodematch inflation' are significant. 
The former, with a positive estimate, suggests that countries with similar educational spending are more likely to trade with each other, while the latter, with a negative estimate, indicates that countries with similar inflation levels are less likely to trade. 
Additionally, in 2020, three other nodal characteristics are significant with positive coefficients, namely, `nodematch agriculture', `nodematch industry', and `nodematch GCF'. This suggests that countries with similar profiles in these areas are also more inclined to trade. These findings reflect increased investment in healthcare and research during the COVID-19 crisis and a shift toward agriculture and industry due to service sector slowdowns.

The nodal characteristic `nodecov unemployment' becomes statistically significant in 2020 with a positive estimate, indicating that the presence of this characteristic increases the likelihood of trade between countries. This finding is consistent with observations during the COVID-19 crisis, where mobility restrictions negatively impacted production and, according to Okun's Law (a direct relationship between production and employment levels; \citealt{okun1962potential}), increased unemployment. This situation may have compelled economies to boost trade to secure essential goods.
Finally, the nodal characteristic `nodefactor landlocked' (access to the sea) reveals a negative and statistically significant relationship with the formation of trade links. This suggests that landlocked countries are less likely to engage in international trade.

\begin{table}[!htb]
\centering
\begin{threeparttable}
\caption{Significance of the estimates: 2018, 2020, and 2022. Significance: p value < 0.1 (.), p value < 0.05 (*), p value < 0.01 (**),  p value < 0.001 (***).}
\label{Significancias_18_20_22}
\small
\begin{tabular}{l|c|c|c}
    \hline
    Variable & 2018 & 2020 & 2022 \\
    \hline                      
    edges & *** &  & *** \\            
    nodecov agriculture & *** & *** & ***\\
    nodecov industry & *** & *** & *** \\
    nodecov foreign investment & *** & *** &  \\
    nodecov PIB per capita & *** & *** & *** \\
    nodecov education expenditure & *** & *** & ** \\
    nodecov GCF & *** & *** & *** \\
    nodecov inflation & *** &  & *** \\
    nodecov political stability & *** & *** & *** \\
    nodecov total population & *** & *** & *** \\
    nodecov services & *** & *** & *** \\
    nodecov unemployment &  & *** & *** \\
    nodematch agriculture &  & * &  \\
    nodematch industry &  & . &  \\
    nodematch foreign investment &  &  &  \\
    nodematch PIB per capita &  &  &  \\
    nodematch education expenditure & * & * &  \\
    nodematch GCF &  & ** &  \\
    nodematch inflation & * & ** & *** \\
    nodematch political stability &  &  &  \\
    nodematch total population &  &  &  \\
    nodematch services &  &  &  \\
    nodematch unemployment &  &  & . \\
    nodefactor landlocked & *** & *** & *** \\
    \hline
\end{tabular}
\begin{tablenotes}
\centering
    \item Source: Prepared by the authors based on data from WITS and the World Bank.
\end{tablenotes}
\end{threeparttable}
\end{table}

\subsection{Stochastic Block Model}

In order to understand the structure of the trade network and cluster countries according to their structural characteristics, we implement here a stochastic block model for each year. The results for all three time periods are nearly identical, both in terms of the number of optimal classes and the resulting groupings. Therefore, we present the results only for the year 2018, as the conclusions and results are consistent across the other periods. This suggests that there was no substantial change in the structure of the international trade network during or after the pandemic.

\begin{figure} [!htb]
    \centering
    \includegraphics[scale=0.40]{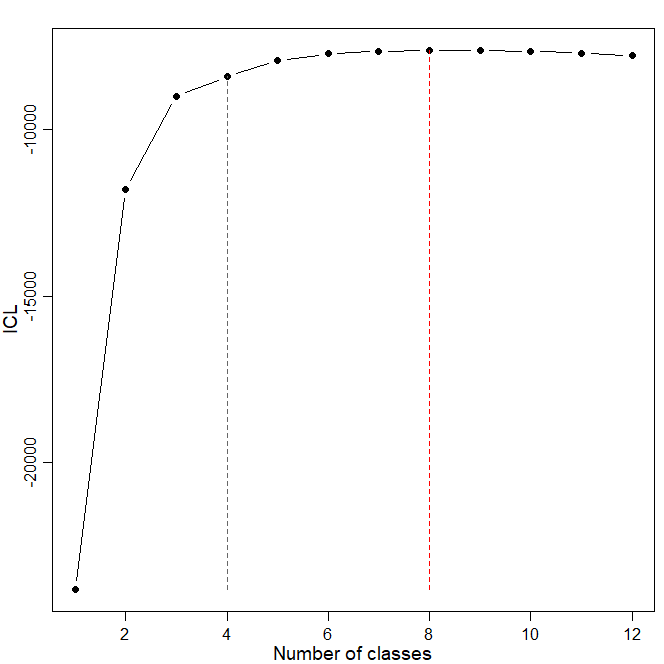}
    \caption{Selection of the number of classes using the ICL criterion.}
    \label{fig:ICL}
\end{figure}

The number of classes and their assignments are typically unknown in real-world data. Here, we use the integrated classification likelihood (ICL) criterion to identify the number of classes. Higher ICL values indicate a better model fit. We test class numbers \( Q \) from 1 to 12. Figure \ref{fig:ICL} shows that the ICL increases with the initial classes and stabilizes at \( Q = 8 \) (see dashed red line). According to \cite{R_analisis_sosa}, while the ICL helps determine the class number, economic meaning favors a lower number of classes. Thus, we select \( Q = 4 \) to maintain interpretability without significantly reducing the ICL.

\begin{figure}[!htb]
    \centering
    \includegraphics[scale=0.85]{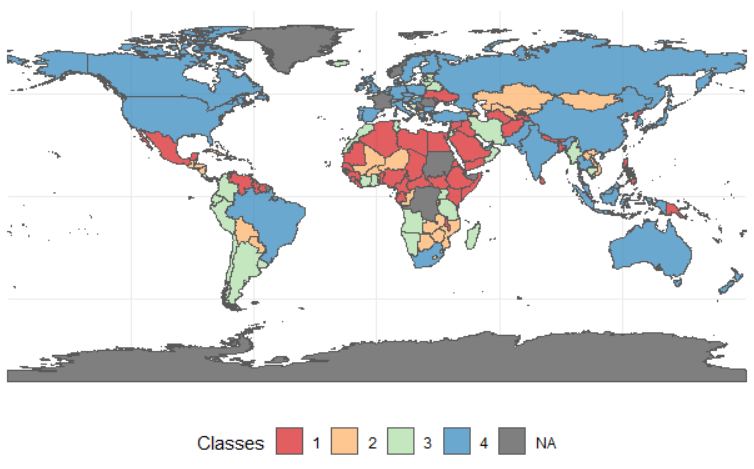}
    \caption{Clustering of countries based on their structural characteristics in the 2018 international trade network, using a SBM.}
    \label{fig:map}
\end{figure}

Now, we fit the SBM using $Q=4$ classes. Results are presented in Figure \ref{fig:map}. Class 1 (79 countries) includes countries such as Vietnam and Azerbaijan; Class 2 (39 countries) includes countries such as Yemen and Palau; Class 3 (39 countries) includes countries such as Slovakia and Chile; and Class 4 (38 countries) includes countries such as China and Germany.
Since the model considers only relational structural characteristics of countries, the resulting clustering reflects commercial development. Generally, countries in Class 1 exhibit less commercial development than those in Class 2, countries in Class 2 exhibit less than those in Class 3, and countries in Class 3 exhibit less than those in Class 4.

\begin{figure}[!htb]
    \centering
    \includegraphics[scale=0.55]{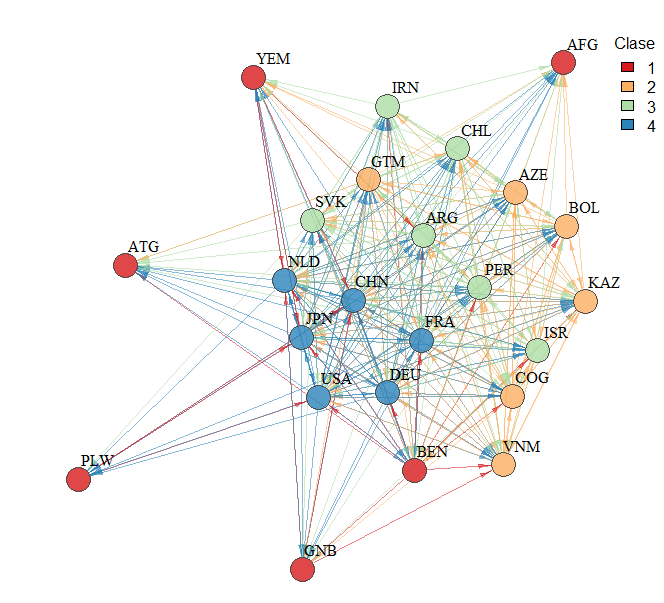}
    \caption{Graph of the international trade network for 2018, differentiated by country classes and filtered to show the six largest exporters in each class. Vertex colors represent country classes, and edge colors correspond to the exporting country.}
    \label{fig:grafo top}
\end{figure}

Figure \ref{fig:grafo top} shows the international trade network segmented those classes using the SBM, focusing on the six countries with the highest export value added per class. 
In the figure, vertex colors represent country classes, and edge colors correspond to the exporting country.
As established in Section \ref{section: Conectividad}, the international trade network includes many countries with low-value connections, while a few generate high-value connections due to their larger economies. 
The predominance of blue edges, linked to the most trade-capable countries, shows that Class 4 countries serve as central hubs. The number of edges decreases for countries in classes 3, 2, and 1, reflecting their lower export capacity.

\begin{figure}[!htb]
    \centering
    \includegraphics[scale=0.55]{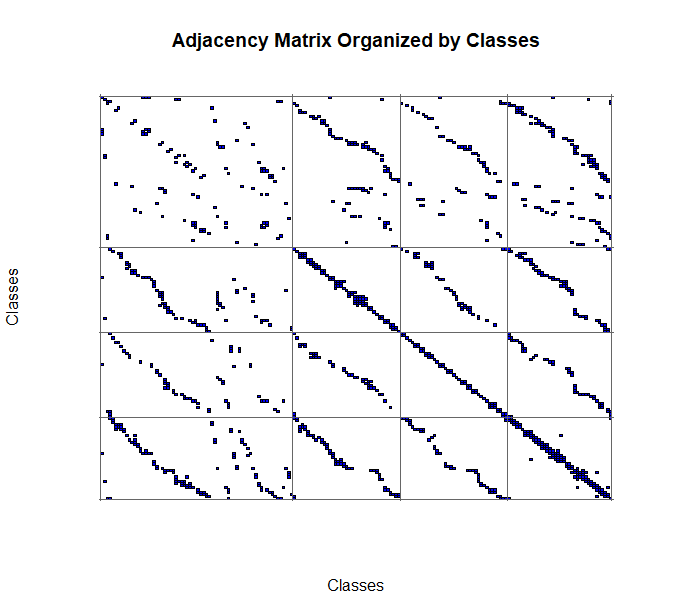}
    \caption{Adjacency matrix ordered by country classes for 2018.}
    \label{fig:matriz_relaciones}
\end{figure}

Figure \ref{fig:matriz_relaciones} presents the adjacency matrix ordered by country classes. There are no clear patterns of relationships within or between classes, which is expected given current trade dynamics. Unlike other networks, where within-class ties are stronger, international trade involves countries trading with both similar and different partners. This is due to the international division of labor, where countries specialize in goods with comparative advantages and trade for those they cannot produce as efficiently.

\section{Discussion}\label{disc}

Both the network structure analysis and the ERGM identify a set of persistent, non-structural node characteristics that are statistically significant and key determinants of international trade for 2018, 2020, and 2022. These determinants include the agriculture, industry, services, GDP per capita, investment, education expenditure, inflation, and access to a coastline. The ERGM results also reveal a shift in significant variables during the pandemic, with unemployment and differences in agricultural, industrial output, and gross capital formation that become relevant. These findings align with those from similar studies, such as \cite{Principal_2022} and \cite{Space_time_2012}.

The SBM divides the network into four classes that reflect countries' export capacities and their importance in the network, distinguishing central hubs from those with less integration. A notable feature of the network is that countries tend to trade with both structurally similar (within their class) and different (from other classes) countries. Lastly, the analysis shows that the global trade network did not undergo significant structural changes during or after the COVID-19 pandemic, as countries' classifications and roles remained largely stable.

The COVID-19 pandemic caused a short-term shock to bilateral trade relationships, weakening transactions in 2020. However, there is no evidence of a long-term structural impact on the determinants of international trade. The pandemic can be seen as a short-term disruptive event that affected the strength of interactions but did not alter the inherent structure of the trade network or the interactions between agents. Therefore, it is expected that globalization dynamics will not be significantly affected in the long term. This is supported by the analysis of the network's structural characteristics, which remained disassortative with an average density around 0.5 and stable connectivity before, during, and after the pandemic.

This document advances understanding of bilateral trade relationships, their determinants, and the international trade network, including the impact of COVID-19. However, it faces limitations. On the one hand, incomplete data for post-2022 periods restricted the analysis of the trade network’s post-pandemic structure. Future research should seek updated data. On the other hand, computational limitations excluded structural network characteristics, like triangles and mutuality, from the ERGM. Including these features in future studies could enhance model accuracy. Such ideas will be discussed elsewhere.

\section*{Statements and Declarations}

The authors declare that they have no known competing financial interests or personal relationships that could have appeared to influence the work reported in this article.

%\nocite{*}
\bibliography{references.bib}
\bibliographystyle{apalike}

%\appendix
%\section{Appendix here}

\end{document}